\documentclass[journal]{IEEEtran}
\usepackage{amsmath,amssymb}
\usepackage[colorlinks=true,bookmarks=false,citecolor=blue,urlcolor=blue]{hyperref}
\usepackage{soul}
\usepackage{amsmath,amssymb,amsfonts}
\usepackage{mathrsfs}
\usepackage{graphicx}
\usepackage{xcolor}
\usepackage{algorithm}
\usepackage[noend]{algpseudocode}
\usepackage{bm}
\usepackage[numbers,sort&compress]{natbib}
\usepackage{comment}
\usepackage{subfig}

\setstcolor{red}

\begin{document}

% \title{WiFi Network Management via LLMs}
\title{WiFi Pathologies Detection using LLMs}
% \author{Forough Shirin Abkenar}
% \thanks{}

\author{\IEEEauthorblockN{Forough Shirin Abkenar~\IEEEmembership{Member,~IEEE}}\\
\IEEEauthorblockA{Ph.D. of Electrical Engineering and Computer Science}% <-this % stops an unwanted space
\thanks{Corresponding author: F. Shirin Abkenar (email: forough.shirin.abkenar@ieee.org).}}

% \author{\IEEEauthorblockN{Forough Shirin Abkenar\IEEEauthorrefmark{1}~\IEEEmembership{Member,~IEEE}}\\
% \IEEEauthorblockA{\IEEEauthorrefmark{1}Ph.D. of Electrical Engineering and Computer Science}% <-this % stops an unwanted space
% \thanks{Corresponding author: F. Shirin Abkenar (email: forough.shirin.abkenar@ieee.org).}}

\maketitle

\begin{abstract}
    In this paper, we fine-tune encoder-only and decoder-only large language models (LLMs) to detect pathologies in IEEE 802.11 networks, commonly known as WiFi. Our approach involves manually crafting prompts followed by fine-tuning. Evaluations show that the sequential model achieves high detection accuracy using labeled data, while the causal model performs equally well for unlabeled data.
\end{abstract}


\section{Introduction}
The wireless network, specifically the wireless local area network (WLAN), broadens the accessibility of the Internet for end users compared to wired connections. Within the array of technologies offered by WLAN, wireless fidelity (WiFi) falls into a broader category which refers to the set of IEEE 802.11 wireless networking standards \cite{wifi_survey}. WiFi provides wireless connectivity to multiple heterogeneous connected devices and the Internet of Things (IoT) devices, including those within highly dense populated areas, such as airports, stadiums, hospitals/healthcare centers, and cafes \cite{Cisco}.

The inherent uncertainties of the wireless medium, resulting from phenomena such as channel fading, interference, limited bandwidth, and propagation environment, amplify the complexity of the wireless medium. With the unprecedented growth of the connections, as reported by Cisco \cite{Cisco}, these challenges are likely to become even more pronounced. Consequently, detecting, localizing, and troubleshooting of pathologies in such an unstable environment pose challenges not only for end users but also for experts.

Research on diagnosing WiFi pathologies has explored various approaches. Some rely on specialized equipment, driver modifications, or application-layer techniques to collect network data \cite{equipment_1,equipment_2}. Others use active probing without modifying 802.11 devices, employing external tools or ML-based models. For instance, Syrigos \textit{et al}. \cite{enabling} developed a framework that detects network issues by monitoring MAC layer statistics, while their later work \cite{ml} used supervised learning with MAC data and modulation schemes to classify pathologies like contention, low SNR, and hidden terminals. Kanuparthy \textit{et al}. \cite{can} introduced WLAN-probe, a user-level tool diagnosing WiFi issues without vendor-specific hardware. Kim \textit{et al}. proposed WiSlow, which distinguishes interference sources (802.11 vs. non-802.11) and estimates the location of non-802.11 interferers using packet loss and transmission rates, applying statistical tests for classification.

Although the aforementioned methods are able to troubleshoot the IEEE 802.11 related pathologies, they suffer from notable limitations that warrant attention. First of all, the methods in the literature tailored to individual users, necessitating adjustments based on each user's specific needs. However, the pathologies are common for all users, suggesting the potential for developing generalized approaches or models that can effectively address these shared issues without the need for extensive customization. Moreover, the above-mentioned methods need to access comprehensive information of the users' devices that rises privacy concerns. Indeed, the critical information, such as detailed device data and network configurations, can be divulged, especially if such information is not essential for resolving WiFi issues. Thus, the existing research work are vulnerable to privacy violation of users information.

Large language models (LLMs) have become integral to AI-driven applications, spanning tasks from translation to text generation, thanks to their transformer-based architecture. LLMs are broadly categorized into encoder-only, decoder-only, and encoder-decoder models, with the first two being most relevant to our work. Encoder-only models, such as BERT, excel in text classification and sentiment analysis, while decoder-only models, such as GPT, generate text by predicting the next word in a sequence \cite{handson}. Our goal is to leverage LLMs for detecting pathologies in WiFi connections. Given network parameters, we first apply prompt engineering to craft effective prompts for pathology detection. To demonstrate LLMs’ efficacy in network applications, we fine-tune an encoder-only model for supervised classification of pathologies. We then extend our approach by fine-tuning a decoder model for unsupervised data, ensuring robust detection across different data scenarios.

\section{IEEE 802.11 Related Pathologies}\label{sec:pathologies}
\begin{figure}[!t]
    \centering
    \includegraphics[width=\columnwidth]{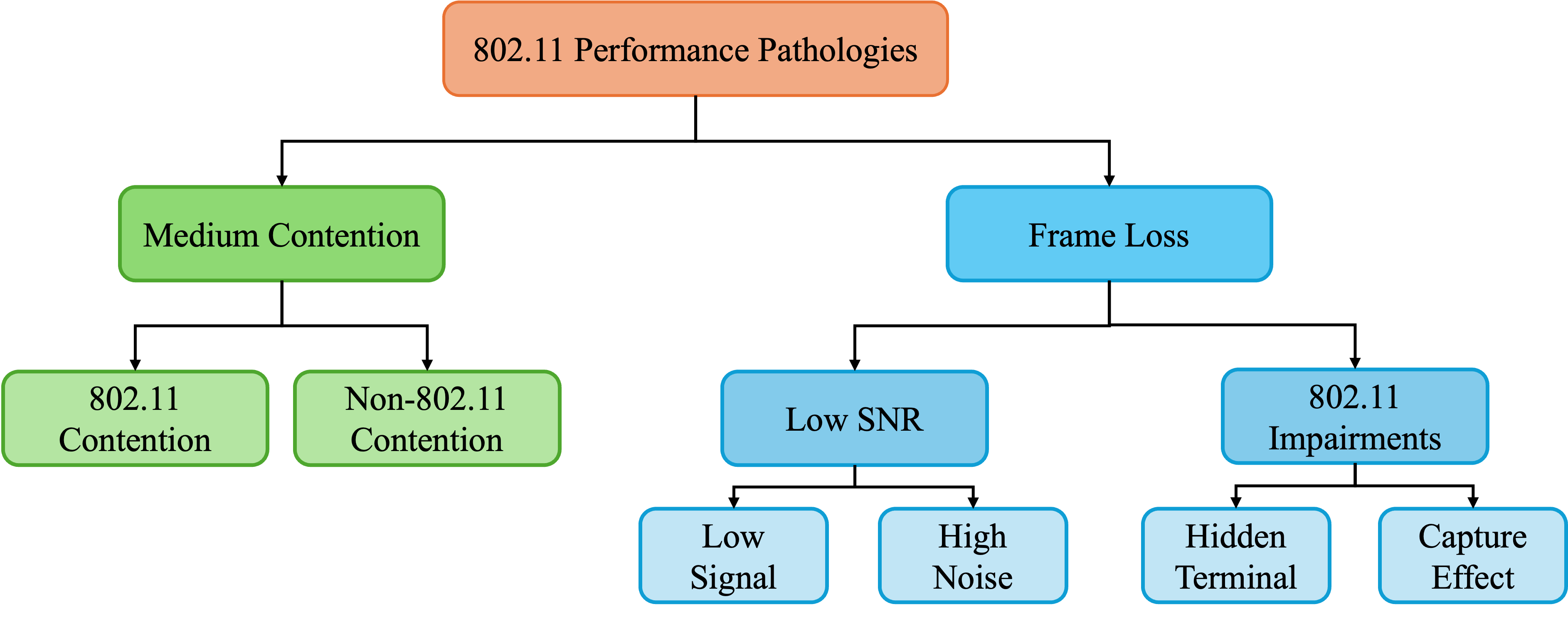}
    \caption{Categories of IEEE 802.11 pathologies \cite{enabling,ml}.}
    \label{fig:pathologies}
\end{figure}

Common pathologies experienced by WiFi networks in real-world scenarios are congestion, contention, hidden terminal, capture effect, low SNR, and interference from non-802.11 devices (such as microwave, cordless phones, and security cameras) \cite{enabling,Cisco_interference}. As shown in Fig. \ref{fig:pathologies}, the authors in \cite{enabling,ml} have divided the 802.11 performance pathologies into two primary categories, i.e., \textit{medium contention} and \textit{frame loss}. The former includes contentions originated from 802.11 and non-802.11 devices. The latter, is categorized into low SNR (due to low signal strength or high noise power) and 802.11 impairments (i.e., hidden terminal and capture effect). In the rest of this section, we elaborate each pathology.

\subsection{Medium Contention}
Medium contention refers to the situation where the transmitter defers the transmission because it recognizes the medium busy. As the name suggests, contention-driven pathologies frequently happen in densely populated areas where numerous devices attempt to concurrently access the wireless medium. Notably, such a contention is not solely adhering to 802.11 standard. Indeed, other protocols (such as Bluetooth and Zigbee), along with radio frequency (RF) devices (such as microwave, cordless phones, and security cameras), also operate within unlicensed spectrum, contributing to increased congestion of the medium. As a result, the main reasons for medium contention are the times that the channel is occupied, also known as channel airtime, by 801.11 transmissions and the duty cycle (DC) of non-802.11 RF devices, i.e., the portion of time that a RF device produces electromagnetic radiation. The former is called  \textit{802.11 Contention} and the latter is named \textit{Non-802.11 Contention} \cite{enabling,wilslow}.

\subsection{Frame Loss}
Unlike the medium contention wherein the transmitter recognizes the channel busy, the transmitter in the frame loss detects the medium as idle. Thus, it initiates the frame transmission. However, the receiver fails to receive the frame successfully. Frame loss happens due to either \textit{Low SNR} or \textit{802.11 Impairments} \cite{enabling,ml}.

\subsubsection{Low SNR}
Signal strength can be attenuated by various phenomena such as channel fading and shadowing, which occur as the signal travels through the wireless medium. These effects can lead to a decrease in the SNR at the receiver, such that if the SNR falls below an acceptable threshold, the receiver may fail to correctly interpret the received signal, resulting in lost or corrupted frames. The other remarkable reason for a low SNR is the high noise power imposed by the non-802.11 devices. Indeed, when the non-802.11 emits high-power electromagnetic radiations, the noise level in the channel rises. Consequently, the probability of the frame loss increases as well \cite{can,ml}.

\subsubsection{802.11 Impairments}
802.11 impairments include phenomena which lead to frame collision. Two most well-known phenomena are \textit{Hidden Terminal} and \textit{capture effect}. Hidden terminal occurs when two transmitters are out of range of each other or obstructed by obstacles so that they cannot directly detect each other. Assuming that both transmitters are in the range of a common access point (AP), if they transmit information frames to the common AP, a potential communication conflict happens and the frames might be collides or lost. On the other hand, the capture effect refers to the situations where two transmitters with different levels of signal strength transmit frames to an AP. In such a situation, the stronger signal overwhelms the weaker signal and capture the receiver for a long time. Thus, frames of the transmitter with weaker signal are lost \cite{enabling,can}.

\section{Dataset and Model Fine-tuning}
\subsection{Dataset}
The dataset used in this study is RadioML 2018.01A, published on Kaggle \cite{Kaggle}. It consists of 24 modulation types, including OOK, ASK4, ASK8, BPSK, QPSK, PSK8, PSK16, PSK32, APSK16, APSK32, APSK64, APSK128, QAM16, QAM32, QAM64, QAM128, QAM256, AM\_SSB\_WC, AM\_SSB\_SC, AM\_DSB\_WC, AM\_DSB\_SC, FM, GMSK, and OQPSK. Each modulation type is recorded at 26 SNR levels ranging from -20 dB to +30 dB in 2 dB increments, with 4,096 frames per modulation-SNR pair. Each frame contains 1,024 complex-valued time-series samples, represented as floating-point in-phase (I) and quadrature (Q) components, resulting in a frame shape of (1024,2). In total, the dataset comprises 2,555,904 frames. This dataset enables the classification of low-SNR conditions under the Frame Loss scenario. Consequently, we define the following labeling policy:

\begin{center}
    \begin{tabular}{|c|c|c|}
        \hline
        \textbf{Pathology} & \textbf{SNR Range} & \textbf{Label}\\
        \hline
        Low Noise & $snr > 15$ & 0\\
        \hline
        Moderate Noise & $5 < snr \leq 15$ & 1\\
        \hline
        High Noise & $-10 < snr \leq 5$ & 2\\
        \hline
        Severe Noise & $snr \leq -10$ & 3\\
        \hline
    \end{tabular}
\end{center}

Relying on the dataset information, we craft customized prompts based on the following template:

"""You are diagnosing WiFi network pathologies based on signal information.\textbackslash nClassify the WiFi condition based on the parameters provided.\textbackslash nParameters: In-phase and quadrature (I/Q) data are \textit{I/Q\_values}. The modulation type is \textit{modulation\_type}. Signal-to-Noise Ratio (SNR) is equal to \textit{snr\_value}.\textbackslash nPathology Type:"""

\subsection{Fine-tuning}
In this section, we fine-tune encoder-only and decoder-only LLMs for supervised and unsupervised data, respectively.

\subsubsection{Fine-tuning an Encoder-only Model}
Encoder-only models like BERT excel in classification tasks. Given that DistilBERT is 40\% smaller and 60\% faster than BERT, we fine-tune DistilBERT to diagnose pathologies in WiFi networks. Table \ref{tab:distilbert_performance} presents the model's performance over three epochs, using a learning rate of 2e-5, a batch size of 32, and a weight decay of 0.01 to mitigate overfitting. We further assess the model on a test dataset with unseen samples. The fine-tuned DistilBERT model achieves an accuracy of 100\% and an F1 score of 1.0.

\begin{table}[!h]
    \centering
    \caption{Performance of Fine-tuned DistilBert model}
    \label{tab:distilbert_performance}
    \begin{tabular}{|c|c|c|c|c|}
        \hline
        \textbf{Epoch} & \textbf{Training Loss} & \textbf{Validation Loss} & \textbf{Accuracy} & \textbf{F1}\\
        \hline
         1 & 0.707 & 0.634 & 0.850 & 0.835\\
         \hline
         2 & 0.098 & 0.076 & 1.00 & 1.00\\
         \hline
         3 & 0.058 & 0.041 & 1.00 & 1.00\\
         \hline
    \end{tabular}
\end{table}

Besides, we employ parameter-efficient fine-tuning (PEFT) and apply low-rank adaptation (LoRA) to minimize computational complexity. LoRA reduces the total parameter count from 67,587,080 to just 630,532 trainable parameters—a 99.07\% reduction—while maintaining strong performance. Table \ref{tab:lora_performance} illustrates LoRA's efficiency in classifying pathologies. Remarkably, the LoRA-based model also achieves an accuracy of 100\% and an F1 score of 1.0, demonstrating a significant reduction in trainable parameters without compromising performance.

\begin{table}[!h]
    \centering
    \caption{Performance of LoRA-based model}
    \label{tab:lora_performance}
    \begin{tabular}{|c|c|c|c|c|}
        \hline
        \textbf{Epoch} & \textbf{Training Loss} & \textbf{Validation Loss} & \textbf{Accuracy} & \textbf{F1}\\
        \hline
         1 & 0.65290 & 0.51726 & 0.93333 & 0.93275\\
         \hline
         2 & 0.00460 & 0.00028 & 1.00000 & 1.00000\\
         \hline
         3 & 0.00060 & 0.00005 & 1.00000 & 1.00000\\
         \hline
    \end{tabular}
\end{table}

% \begin{figure}[t!]
%     \centering
%     \includegraphics[width=5in]{WiFi_Network_Management/images/lora_performance.png}
%     \caption{Performance of Fine-tuned LoRA model.}
%     \label{fig:lora_performance}
% \end{figure}

\subsubsection{Fine-tuning a Decoder-only Model}
We repeat the experiments considering unsupervised data. Hence, we fine-tune the GPT-2 model. Table \ref{tab:gpt_performance} presents the performance of the model over two epochs, using the default learning rate of 5e-5, a batch size of 4, and a weight decay of 0.01 to mitigate overfitting.

\begin{table}[!h]
    \centering
    \caption{Performance of Fine-tuned GPT-2 model}
    \label{tab:gpt_performance}
    \begin{tabular}{|c|c|c|}
        \hline
        \textbf{Epoch} & \textbf{Training Loss} & \textbf{Validation Loss}\\
        \hline
         1 & 0.4769 & 0.4648\\
         \hline
         2 & 0.4608 & 0.4564\\
         \hline
    \end{tabular}
\end{table}

\begin{figure}
     \centering
     \subfloat[][High Noise]{\includegraphics[width=\columnwidth]{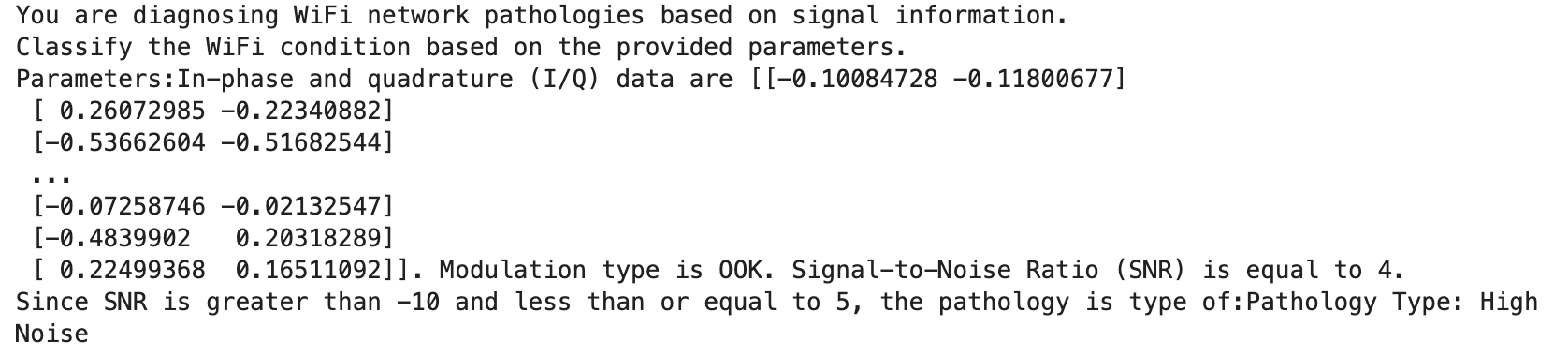}\label{fig:high_noise}}\\
     \subfloat[][Low Noise]{\includegraphics[width=\columnwidth]{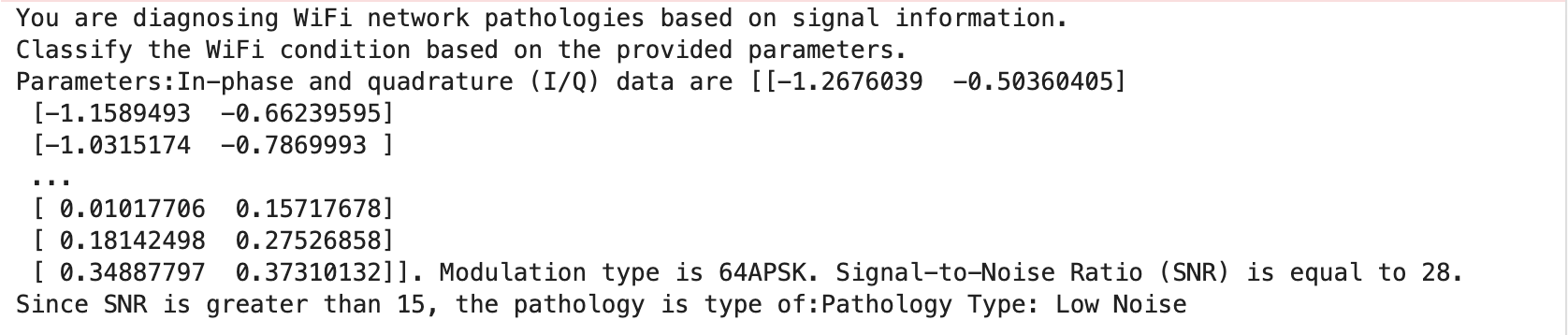}\label{fig:low_noise}}\\
     \subfloat[][Severe Noise]{\includegraphics[width=\columnwidth]{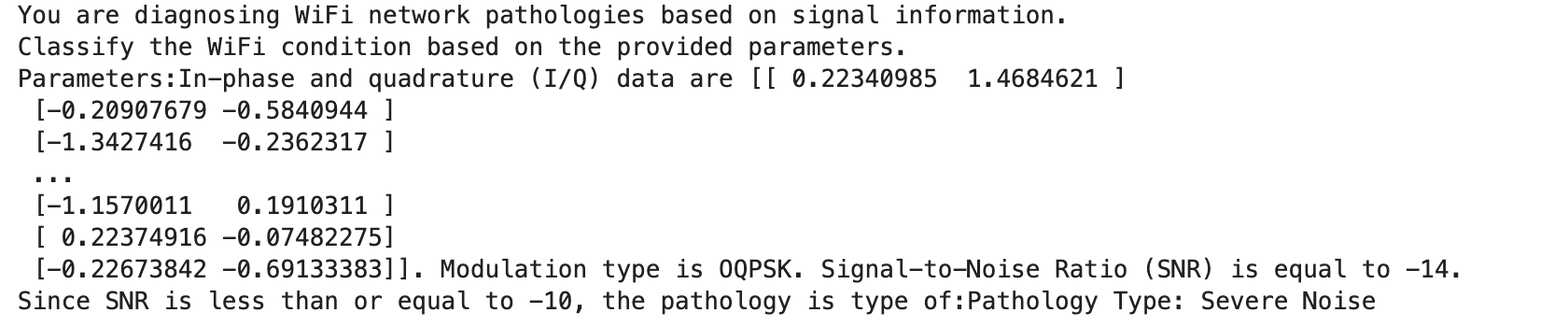}\label{fig:severe_noise}}
     \caption{Performance of the fine-tuned causal model.}
     \label{fig:causal}
\end{figure}

We further assess the model on a test dataset with unseen samples. Figure \ref{fig:causal} shows examples of pathologies detected by the fine-tuned causal model. The obtained results reveal that the fine-tuned model is successful in detecting corresponding noise-driven pathologies.

\section{Conclusion}
\noindent In this paper, we demonstrated the effectiveness of LLMs in wireless communications for detecting WiFi pathologies. We fine-tuned encoder-only and decoder-only LLMs for supervised and unsupervised data, respectively, using manually crafted prompts tailored for pathology detection. Due to limited available data, our study focused on noise-driven pathologies, though existing research highlights additional causes, i.e., signal strength, hidden terminals, and capture effects. In future work, we aim to generalize our model to encompass all pathology types and enhance it to not only detect but also troubleshoot network issues.

\bibliographystyle{IEEEtran}
\bibliography{main}

\end{document}